\documentclass[aps,prd,preprint,floats,tightenlines]{revtex4}

\newcommand{\beq}{\begin{equation}}
\newcommand{\eeq}{\end{equation}}

\begin{document}

\title{Perspectives in cosmology}

\author{Alexander Vilenkin}

\address{
Institute of Cosmology, Department of Physics and Astronomy,\\ 
Tufts University, Medford, MA 02155, USA
}


\begin{abstract}

The ``new standard cosmology", based on the theory of inflation, has
very impressive observational support.  I review some outstanding
problems of the new cosmology and the global view of the universe --
the multiverse -- that it suggests.  I focus in particular on
prospects for further observational tests of inflation and of the
multiverse.

\end{abstract}

\maketitle

\section{Introduction}

We live at a very interesting juncture in cosmology, when a new
paradigm appears to be shaping up.  Its true significance will only be
clear from the perspective of the future.  In this talk I will try to
outline where I think we stand and where we might be
going.\footnote{Limitations of space do not allow me to provide
  adequate references to the literature, so I chose to give no
  references at all.}  This is, of course, a tricky business.  When
speaking about perspectives, it's good to remember the quote from
George Orwell, who wrote that even ``to see what is in front of one's
nose needs a constant struggle."

There are three basic questions in cosmology that we would like to
answer.  The first is: What is out there?, or What is the content of
the universe?  The second is: How did it get there?, or How did the
universe evolve to its present state? And finally, What does it all
mean?  What is the big picture and how is it related to fundamental
physics?  There have been remarkable developments in all three of
these areas over the last few decades.

The picture of the universe which has emerged from these developments
can be called ``the new standard cosmology."  According to this
picture, the universe has a nearly flat geometry, and its main three
ingredients are dark energy (72\%), dark matter (23\%), and baryonic
matter (less than 5\%).  The key role in the new cosmology is played
by inflation -- a period of super-fast, accelerated expansion in the
early universe.  Density perturbations that led to structure formation
originated as quantum fluctuations during this inflationary period.
Baryonic matter was formed in non-equilibrium, baryon-number-violating
processes after inflation.  After that, the evolution is along the
lines of the hot big bang model.  The new cosmology is still under
construction, with details remaining largely uncertain.  I will
discuss some of these details in the rest of the talk.

\section{What is out there?}

With some luck, we may be on the verge of discovering what dark matter
is.  Some of the dark matter candidates are well motivated from the
particle physics perspective.  For example, weakly interacting massive
particles (WIMPs) could be produced as thermal relics in the early
universe and are well suited for direct and indirect searches and for
collider experiments.  Axions could be produced by vacuum misalignment
during inflation.  There is no shortage of other candidates:
supermassive particles, gravitinos, Q-balls, vortons, etc.  I will
stop at that, since dark matter will be discussed extensively at this
meeting.

Turning now to dark energy, in Frank Wilczek's words, it is ``the most mysterious fact in all of physical science."  We have several dark energy candidates, but all of them have serious problems.  First, the dark energy may simply be the vacuum energy, or the cosmological constant, $\rho_v$.  The vacuum should then have pressure $P_v=-\rho_v$, with both $P_v$ and $\rho_v$ constant in time.  A widely discussed alternative is a dynamical dark energy.  Phenomenologically, it can be characterized by an equation of state, 
\beq
P=w\rho,
\eeq
where $w\neq -1$ and may be time-dependent.  In quintessence models, the dark energy is attributed to a scalar field with a slowly declining potential, e.g., 
\beq
V(\phi)\propto \phi^{-\alpha}.  
\eeq
As the field rolls down the potential, its energy density gradually gets smaller.  The dark energy is so small today because the universe is so old, and the field $\phi$ had time to roll to very large values.  Finally, there are modified gravity models, which explain the observed accelerated expansion by deviations from Einstein's gravity.  Of all three alternatives, I think the front runner is the cosmological constant, for the reasons that I will now explain.

As I have mentioned, the cosmological constant has its problems.  The first of these is the old cosmological constant problem, which has been with us for nearly half a century.  All quantum fields, like the electromagnetic field, contribute to the vacuum energy.  This contribution is given by the sum of zero-point oscillation energies over modes with all possible frequencies.  This is a divergent sum, and we should cut it off at some mass scale $M$, which may be set, for example, by the supersymmetry breaking scale,
\beq
\rho_v = {1\over{V}}\sum{1\over{2}} \omega \sim M^4.
\eeq
(I use the system of units where $\hbar=c=1$.)
With $M\sim 1$ TeV, this gives $\rho_v\sim 10^{60}\rho_m$, where $\rho_m$ is the average density of matter, while observations indicate that $\rho_v\sim 3\rho_m$.  Fermions and bosons contribute to $\rho_v$ with opposite signs, but we would need cancellation up to the 60th decimal point in order to explain the observed value.

Another problem is related to the fact that the vacuum energy density remains constant in time, while the matter density is diluted by the expansion.  Why then do we happen to live at the very special epoch when the two densities are comparable?  This is the so-called coincidence problem.

As of now, the only viable resolution of these problems is provided by the anthropic approach.  It assumes that $\rho_v$ takes a wide range of values in different parts of the universe.  Of course, this can work only if the underlying particle physics admits many vacua with different values of $\rho_v$.  This picture has some support from string theory, and I will get back to it toward the end of the talk.  For now, let us assume that $\rho_v$ is indeed variable from one region to another.

The vacuum energy acts as a repulsive force, and it prevents galaxies
from forming once it comes to dominate the universe.  So, a large
$\rho_v$ means no galaxies.  We can calculate the probability
distribution for an observer randomly picked in the universe to
measure a given value of $\rho_v$.  In regions with large values of
$\rho_v$ there are no galaxies and therefore no observers, so the
probability is strongly suppressed.  Very small values, where $\rho_v$
is much smaller than it needs to be for galaxy formation, are also
rather unlikely.  One finds that the observed value of $\rho_v$ is
well within the $2\sigma$ range of the resulting distribution.  It is
worth pointing out that the anthropic prediction of $\rho_v \sim
\rho_m$ was made well before the observation.\footnote{I would like to
  emphasize that this prediction is not based on the trivial
  statement, often called ``the anthropic principle", that observers
  can live only in regions where the constants of nature are
  consistent with their existence.  Rather, it is based on ``the
  principle of mediocrity", which asserts that we are typical
  representatives in a class of similar observers in the universe.
  Although not always explicitly stated, this principle is always
  assumed in statistical predictions, such as the prediction of the
  multipoles of the CMB.}

The main drawback of the alternative dynamical dark energy models is that they do not address the cosmological constant problems.   For example, quintessence models assume that the dark energy density approaches zero as $\phi\to\infty$, while in general it should approach a nonzero vacuum energy density $\rho_v$.  These models also offer no solution to the coincidence problem.  The same criticisms apply to the modified gravity approach.  Moreover, the equation of state parameter $w$ derived from the WMAP data combined with supernova observations is
\beq
w=-0.97\pm 0.07,
\eeq
which is very close to the cosmological constant value $w=-1$.  

Apart from the three main components of the universe, what else may be out there?  We know there are ultrahigh-energy cosmic rays.  Until recently their origin appeared mysterious, but now it seems that they may have a rather mundane astrophysical origin.

There is much to be learned about cosmic magnetic fields.  We have a rather sketchy information about the field distribution on the largest scales, and the origin of the magnetic fields remains a mystery.

We believe there should be gravitational waves, and we expect to detect them in the coming decade.

There are also some more exotic things which may or may not be there.  For example, topological defects, such as cosmic strings.

Of course, there may also be some surprises -- things we have not thought about which will change the direction of current research in new and unexpected ways.   

I would like to discuss cosmic strings in a bit more detail.  Strings are linear defects which could be formed at a phase transition in the early universe.  They are predicted in a wide class of elementary particle models.  Some superstring-inspired models suggest that fundamental strings may also have astronomical dimensions and play the role of cosmic strings.  

Once they are formed, strings evolve in a scale-invariant manner.  A
typical horizon volume at any time has some long strings stretching
across it and a large number of small closed loops.  The average
number of long strings per horizon is\footnote{Both the normalization
  and the $p$-dependence in this equation are somewhat uncertain.  The
  dynamics of a cosmic string network involves highly non-linear
  processes on a wide range of length scales; as a result, a
  quantitative description of string evolution is an extremely
  difficult problem.  This problem is now successfully being addressed
  in new, high-resolution numerical simulations of cosmic strings.}
\beq 
N\sim 10/p, 
\eeq 
where $p$ is the probability for two intersecting strings to
reconnect.  Field-theory, solitonic strings always reconnect, so
$p=1$, while for superstrings $p\ll 1$.  This opens an exciting
possibility of directly testing superstring theory.  If we detect
$N\gg 1$ long strings in the sky, we will know that we are looking at
superstrings.

Strings can be detected through a variety of observational effects.
Oscillating loops of string emit gravitational waves -- both bursts
and a stochastic background.  They can also be sources of
ultrahigh-energy cosmic rays.  Long strings can act as gravitational
lenses and can produce characteristic signatures in the CMB:
discontinuous temperature jumps along the strings and a B-mode
polarization pattern due to rotational perturbations induced by
strings.  If cosmic strings are discovered, this would open a window
into physics of very high energies, which are not likely ever to be
accessed in accelerators.

\section{Cosmic history}

There is a great deal yet to be learned about the structure formation
in the universe, from the dark ages, through star and galaxy
formation, and the formation of supermassive black holes.  We are now
in the era of large surveys and massive computer simulations, and we
can expect a lot of progress in this area in the coming decade.  Here,
I will focus on the theory of inflation and its upcoming tests.

A period of inflation in the early universe explains why the universe
is large, expanding, homogeneous, isotropic, and flat.  It is
remarkable that no viable alternatives to inflation have been
suggested, despite some heroic attempts.  This is paralleled only by
Darwin's theory in biology.  And like Darwin's theory, inflation is
not a specific model, but rather a paradigm encompassing a wide class
of models.

Simple models of inflation make very definite predictions:  (i) a flat universe, with the total density very close to the critical density,
\beq
\Omega_{total}=1\pm 10^{-5};
\eeq
(ii) a nearly flat spectrum of adiabatic, Gaussian density fluctuations,
\beq
P_s(k)\propto k^{-n_s},
\eeq
with the spectral index in the range
\beq
0.95 < n_s < 0.98;
\eeq
(iii) a stochastic background of gravitational waves with a flat spectrum.  The amplitude of these gravitational waves is model-dependent, but in the simplest model with a quadratic inflaton potential, 
\beq
V(\phi)= {1\over{2}}m^2\phi^2, 
\label{V}
\eeq
the ratio of the power in tensor and scalar perturbations is
\beq
r\equiv P_t/P_s \approx 0.14.
\eeq 
A gravitational wave background of this magnitude may be within the reach of Planck satellite.  Current observations yield the values
\beq
\Omega_{total}=1\pm 0.01, ~~~~~~ n_s=0.96\pm 0.01, ~~~~~~ r<0.22,
\eeq
and are consistent with the simplest model.  

If the simplest quadratic model (\ref{V}) is supported by subsequent tests, this would be a spectacular triumph of inflationary cosmology.  On the other hand, the CMB data show some tantalizing hints of non-Gaussianity.  The parameter $f_{NL}$, which is related to the three-point correlation function and should vanish for Gaussian fluctuations, is
\beq
f_{NL}=30\pm 15.
\eeq
This is consistent with the Gaussian value $f_{NL}=0$ at the $2\sigma$ level, but barely.  If indeed $f_{NL}\sim 30$, then Planck will detect it at $6\sigma$ level.
That would be a very dramatic development, as it would rule out a wide class of models of inflation, including most single-field models.

Moreover, a number of anomalies have been reported in the CMB data:
(i) the CMB quadrupole is much smaller than expected; (ii) a strange
alignment between the low multipoles; (iii) an asymmetry between the
correlation functions in the Northern and Southern hemispheres; (iv) a
statistically unlikely cold spot in the WMAP microwave map of the sky.
Are these anomalies due to chance?  If you look hard enough at any
stochastic pattern, you will start seeing things.  For example, you
can see my initials ``AV" in the left upper part of the
map.\footnote{I am grateful to Andrei Linde for pointing this out.}
But it is hard to believe that all anomalies are due to chance, so
some new physics may well be forthcoming from all this.

\section{The big picture}

Inflationary cosmology has profound implications for the global
structure of the universe, particularly for what the universe is like
beyond the horizon, and for the beginning and the end of the universe.

The key point is that inflation is generically eternal.  Even though
it has ended in our local neighborhood, it still continues in remote
parts of the universe.  The reason is that the end of inflation
involves quantum, probabilistic processes and does not occur
everywhere at once.  The details are somewhat model-dependent.  For
example, in models where inflation ends through bubble nucleation, we
have an inflating high-energy false vacuum.  Low-energy bubbles
nucleate and expand in this inflating background.  We live in one of
the bubbles and can see only a small part of it.  Even though the
bubbles expand at speeds close to the speed of light, they rarely
collide, because the false vacuum that separates them is expanding
even faster.  This makes room for more bubbles to form, and the
process continues forever.

The physical properties of all low-energy bubbles do not have to be
the same.  In fact, string theory, which is at present our best
candidate for the fundamental theory, appears to have a multitude of
solutions describing vacua with different values of the low-energy
constants of nature.  These solutions are characterized by different
compactifications of extra dimensions, by branes wrapped around extra
dimensions is different ways, by different values of the fluxes, etc.
The number of possibilities is combinatorial and can be as high as
$10^{1000}$.  Bubbles of all possible vacua will nucleate in the
course of eternal inflation.  The resulting multiverse, with bubbles
within bubbles within bubbles, provides a natural arena for anthropic
applications, such as the one I described for the cosmological
constant.

The multiverse picture makes many people rather unhappy.  There have
been complaints that it cannot be tested, even in principle.  If only
a small part of one bubble is accessible to observation, how can we
ever verify that other bubbles really exist?

I think such a test may in fact be possible.  We can look for
observational signatures of bubble collisions.  Even though collisions
are rare, our bubble will collide with an infinite number of other
bubbles in the course of its expansion.  A bubble collision could
produce an imprint in the CMB; its detection would provide direct
evidence for eternal inflation.  There is, however, no guarantee that
a bubble collision has occurred within our cosmic horizon.

Another approach is to use theoretical models of the multiverse to
make statistical predictions for what we can expect to see in our
observable part of the universe.  One such prediction, for the
cosmological constant, has already been confirmed.  This may be our
first evidence that there is indeed a huge multiverse out there.  We
can also derive predictions for other parameters, such as the density
parameter $\Omega$, the ratio of baryonic and dark matter densities,
the neutrino masses, etc.  With more successful predictions, we may be
able to prove the case for the multiverse beyond reasonable
doubt.\footnote{A long-standing problem with making predictions in the
  multiverse stems from the fact that any event having a nonzero
  probability will happen in the course of eternal inflation, and it
  will happen an infinite number of times.  Statistical predictions
  are based on relative frequencies of events in the limit of
  $t\to\infty$.  One finds however that the outcome sensitively
  depends on the limiting procedure.  In the last few years there has
  been a significant progress towards resolution of this problem
  (which is known as ``the measure problem").}

If inflation has no end, could it also have no beginning?  This would
allow us to avoid many perplexing questions associated with the
beginning of the universe.  However, a theorem that I proved with
Arvind Borde and Alan Guth states that, under very general conditions,
inflationary spacetimes are incomplete to the past.  This indicates
that inflation must have had some sort of a beginning.

What could have happened before inflation?  In my (biased) view, the
most attractive possibility is the spontaneous nucleation of a small
closed universe out of nothing.  The nucleation process can be
elegantly described in terms of a Euclidean instanton solution of
Einstein's equations.  Going beyond this semiclassical description
will require further progress in quantum gravity.

I would like to conclude with some important news about the end of the
universe.  It is often said that if the dark energy is a cosmological
constant, then the universe will expand forever.  This is true for the
universe as a whole, but not necessarily for our local region.  In the
multiverse picture, there must be a large number of negative-energy
vacua, and bubbles of such vacua will inevitably form within our
(nearly zero-energy) vacuum.  At some point, probably in a very
distant future, our neighborhood will be engulfed by a negative-energy
bubble.  The expansion will then locally turn into contraction, and
our region will collapse to a big crunch.  The good news is that the
bubble will arrive at nearly the speed of light, so we will have no
time to worry about it.

\end{document}